\documentclass[aip,graphicx]{revtex4-1}

\draft 
\usepackage{color}
\usepackage{graphicx}
\begin{document}


\title{Fiber-coupled pillar array as a highly pure and stable single-photon source} 



\author{S.~Odashima}
\email[]{s-odashima@hi-tech.ac.jp}
\affiliation{Department of Mechanical Engineering, Hachinohe Institute of Technology, Hachinohe 031-8501, Japan}
\author{H.~Sasakura}
\affiliation{Division of Applied Physics, Hokkaido University, Sapporo 060-8628, Japan}
\author{H.~Nakajima}
\thanks{Present address: CNT-Application Research Center, National Institute of Advanced Industrial Science and Technology,
Tsukuba 305-8565, Japan}
\affiliation{Research Institute for Electronic Science, Hokkaido University, Sapporo 001-0021, Japan}
\author{H.~Kumano}
\affiliation{Research Institute for Electronic Science, Hokkaido University, Sapporo 001-0021, Japan}
\affiliation{College of Creative Studies, Niigata University, Niigata 950-2181, Japan}


\date{\today}

\begin{abstract}
A highly pure and stable single-photon source is prepared that comprises a well-designed pillar array
in which each pillar contains only a few InAs quantum dots. A nano-pillar in this array
is in direct contact with a fiber end surface and cooled in a liquid-He bath.
Auto-correlation measurement shows that this source provides an average $g^{(2)}(0)$ value of
0.0174 in the measured excitation-power range.
This photon source and fiber coupling are quite rigid against external disturbances
such as cooling-heating cycles and vibration, with long-term stability.
\end{abstract}


\maketitle 

\section{Introduction}

The generation of a single photon and its on-demand operation provides highly secure information technology
based on quantum cryptography.~\cite{RMP74_145}
In general, single-photons are provided by optical transitions between discrete energy levels
in which the occupation number is limited by the basic principle of quantum mechanics.
This phenomenon is realized in single atoms,~\cite{PRL39_691, PRL89_067901, Science309_454}
molecules,~\cite{PRL69_1516, Nature407_491} and ions~\cite{PRL58_203, Nature431_1075}
(in which the quantized internal energy is inherent because of their size-scale nature)
and in the color centers of diamonds~\cite{PRL85_290, PRL89_187901, NL11_198, NatNanotechnol5_195, NatPhoton6_299}
and semiconductor quantum dots (QDs)~\cite{PRB63_121312(R), PRL86_1502, Finley01, APL88_081905,
Munsch09, NatPhoton4_174, SciRep5_14383, APL101_161107, APL90_061103, APL93_021124, APEX6_065203,
APEX8_112002, APL108_011112, APL110_142104, APL99_121101, SR5_14309, APEX9_032801, APEX6_062801,APL102_131114,
JAP116_043103, APEX1_061202, JAP104_083508}
(which are localized and energetically isolated in the bulk system). Among these single-photon sources (SPSs),
QDs are quite promising as they are realistically applicable.~\cite{QD-BOOK_01, QD-BOOK_03}
This is because the QD density and photon energy are tunable according to our application purposes,
and present semiconductor technologies are available for fabricating suitable optical devices. 
To evaluate {\it single}-photon nature, Hanbury Brown and Twiss (HBT)~\cite{Nature177_27}-type measurements
are usually performed.~\cite{PRL86_1502, APL88_081905, NatPhoton4_174, SciRep5_14383}
The second-order correlation function at zero time delay, $g^{(2)}(0)$,
indicates the purity of a single photon.
By using highly pure single-photon emission, long-distance quantum-key distribution (QKD) has been realized
on an applicable level for a practical telecom QKD network.~\cite{SciRep5_14383}
In many cases, the abovementioned HBT and/or QKD experiments are performed on the free-space optical setup.
Therefore, quite delicate treatment is required to sustain the microscopically optimized optical alignment
during experiments against external disturbances. 

On the other hand, direct contact of SPSs to a fiber end surface~\cite{NL11_198, APL90_061103,
APL93_021124, APEX6_065203, APEX8_112002, APL108_011112, APL110_142104} has also been examined because of the guarantee
of a simple, stable optical coupling. In our previous work,~\cite{APEX6_065203, APEX8_112002}
a semiconductor flake containing InAs QDs was sandwiched by two single-mode fibers (SMFs).
It had a mechanically solid structure, and once it was set in the liquid-He vessel, it worked
as a stable photon source until the liquid He was exhausted.
Although it had a simple structure and worked as a robust photon source, it was difficult to separate
the aimed single-photon signal from background photoluminescence (PL)
because any number of QDs could couple to the fiber core and contribute to the PL spectrum. 
Therefore, controlling the number of QDs that couple to the fiber core is a major issue faced for the development
of this device as an SPS.
Along this direction, low density QD samples~\cite{APEX1_061202, JAP104_083508} and/or
semiconductor processing have been applied to reduce the available QD number.

Here we fabricate a semiconductor pillar array in which each pillar contains only a few QDs.
This pillar array is mounted on the sample stage and directly coupled to the SMF end surface. 
The array structure is well designed as only one or (accidentally) two pillars can couple to the fiber core
without any precise manipulation. We demonstrate that it works well as an SPS and is quite robust
against external disturbances such as vibration and heat cycles.
Our results will provide a key to develop the QD-based single-photon emitter into a practical device.

\section{Sample preparation and experiments}

We use semiconductor QDs as SPSs. InAs QDs are grown on a GaAs (001) substrate by molecular-beam epitaxy (RIBER, MBE32P).
After the growth of 300 nm GaAs buffer layer at 625 $\mathrm{^{\circ}C}$, InAs QDs are grown
with a growth rate of $1.5 \times 10^{-3}$ ML/s at 475 $\mathrm{^{\circ}C}$, then covered by 50 nm GaAs at 600 $\mathrm{^{\circ}C}$.
A QD density is estiamted at about $7 \times 10^{9}$ /$\mathrm{cm^2}$.
A pillar array structure is fabricated using electron-beam lithography (ELIONIX, ELS-F125-U) and reactive-ion etching
(SAMCO, RIE-101iHS). Pillars have a diameter of 300 nm, and the array has a square lattice structure with a distance
of 2.5 $\mathrm{\mu m}$ between pillars (Fig.1(a)). This sample is spin-coated by HSQ (Dow Corning Toray, Fox(R) 15 Flowable oxide)
to protect against mechanical damage (Fig.1(b)), placed in direct contact with the ferrule of an SMF
(Thorlabs, UHNA3, NA = 0.35, mode field diameter (MFD) = 2.6 $\mathrm{\mu m}$ at 1,100 nm),
and tightly fixed by pushing it with another SMF from the back of the sample.
It is set in a liquid-He vessel and cooled at 4.2 K (Fig.1(c)).
Considering the distance between pillars and the mode diameter of a fiber, it can be seen that only one or two pillars
can couple to a fiber core.
Each pillar contains less than 10 QDs in the case of a QD density of $\mathrm{10^{9} \sim 10^{10} \ /cm^{2}}$
and a pillar diameter of 300 nm. QDs located near the edge of a pillar are usually optically inactive.
Therefore, only a few QDs located near the center of a pillar can contribute to luminescence.
Whether this structure is successful or not strongly depends on the abovementioned geometric relation
of a pillar diameter and QD density. In the case that the pillar diameter is well controlled considering QD density
and the outer inactive area in a pillar, each pillar in an array works as an SPS with a high probability.
The area of a pillar array, 500 $\mathrm{\mu m}$ square ($201 \times 201$ pillars) is large enough
to couple this array to the fiber core by eye without any precise manipulation.
The relationship between MFD (2.6 $\mathrm{\mu m}$) and the lattice constant of a pillar array (2.5 $\mathrm{\mu m}$)
guarantees that one of the pillars (but undefined) in the array always couples to the fiber core.

\begin{figure}[t]
\begin{center}
\includegraphics[width=20.5pc]{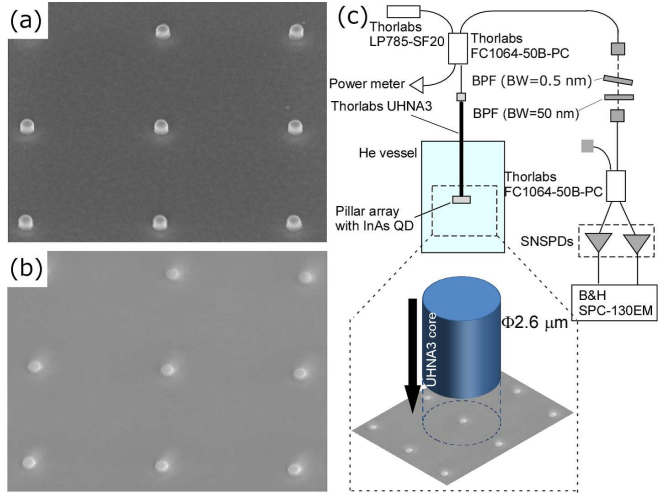}\hspace{2pc}%
\caption{\label{fig1} (a) Pillar array structure with a diameter of 300 nm and a distance between pillars of 2.5 $\mathrm{\mu m}$.
(b) The pillar array is coated with HSQ for protection against mechanical damage. (c) Experimental setup for auto-correlation measurement.
The pillar array is directly coupled to an SMF (MFD = 2.6 $\mathrm{\mu m}$). }
\end {center}
\end{figure}

\begin{figure}[t]
\begin{center}
\includegraphics[width=20.5pc]{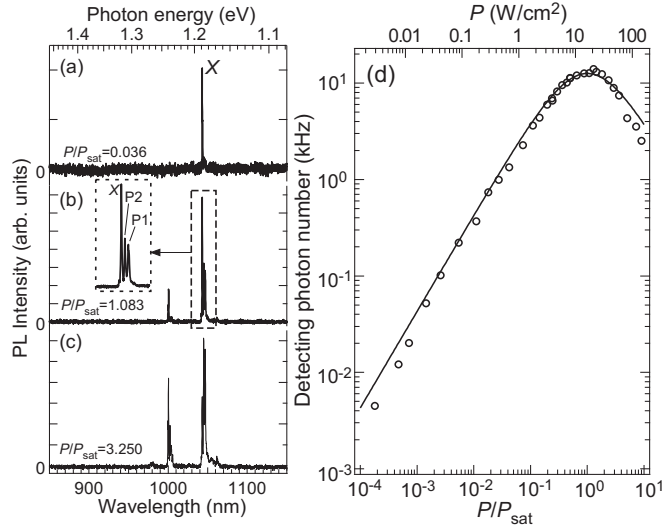}\hspace{2pc}%
\caption{\label{fig2} Typical PL spectra under cw-excitation at $P/P_{sat}$ values of 0.036 (a), 1.083 (b), and 3.250 (c).
(d) $P$ dependence of the detected photon number of the $X$ line (solid open circles). The solid line is fitted by Eq. (\ref{eq1}).
The parameters in Eq. (\ref{eq1}), $\tau _{\textrm {x}}=1.60$ ns and $\tau _{\textrm{xx}}=0.72$ ns are estimated by
the transient PL of the exciton and biexciton states, respectively. $N_{0}=2.92 \times 10^{1}\pm 0.5$ [kHz]
and $\alpha=0.057 \pm 0.002$ [GHz cm$^2$/W] are fitting parameters.
 }
\end {center}
\end{figure}

For PL measurements, we use a fiber-pigtailed laser diode (Thorlabs: LP785-SF20) that emits a laser beam at 785 nm
as an excitation source arriving at the pillar array. To clean up laser spectral noise, a bandpass filter is inserted
into the laser-beam path (Edmund Opt.: \#68-947).  
To spatially separate emission in the reflection direction, a fiber-based  beamsplitter module (Thorlabs: FC1064-50B-PC) is used.
To check the time-integrated PL spectrum, the emission is dispersed by a double-grating spectrometer
(Acton: Spectrapro 2500i, $f=1.0$ m), and photons of each emission energy are detected by a liquid-nitrogen-cooled
InGaAs photodiode array (Roper: OMA-V1024).

InAs QDs are non-resonantly excited with a continuous-wave (cw) condition through the SMF.
Figures~\ref{fig2}(a)-(c) show the time-integrated PL spectra measured on a pillar array under typical excitation
power conditions. As shown in Fig.~\ref{fig2}(a), the well-resolved single peak centered at 1,043.1 nm is observed
with a 0.6-meV full width at half maximum at low excitation power of $P/P_{sat}=0.036$.
$P/P_{sat}$ is the normalized excitation power, where $P$ is the excitation power density estimated from the monitored power
and MFD of the SMF.
$P_{sat}$ is the excitation power at which the detected photon number becomes saturated.
Figure~\ref{fig2}(d) shows the $P$ dependence of the photon count rate $N$ at this peak.
It is deduced that this peak is attributed to the exciton $X$ from the linear behavior of $N$ with $P$ in the weak excitation range.
This $X$ line is selected through a 0.5-nm-wide bandpass filter (Optoquest: custom-made product)
and recorded by a gated photon counter (Stanford Research System Inc.: SR400) with a superconducting
nanowire single photon detector (SNSPD, Single Quantum: custom-made product).
$N(P)$ and $P_{sat}$ are estimated as
\begin {equation}
N(P)=N_{0}\left (1+\frac {1}{\alpha P \tau _{\textrm{x}}}+\alpha P \tau _{\textrm {xx}} \right )^{-1} \label{eq1}
\end {equation}
and
\begin{equation}
P_{sat}=\left(\alpha \sqrt{\tau _{\textrm{x}}\tau _{\textrm{xx}}}\right)^{-1},\label{eq2}
\end{equation}
respectively.
$P_{sat}$ is derived from the condition that $N(P)$ is maximal.
$N_{0}$ ($=2.92 \times 10^{1}\pm 0.5$ [kHz]) and $\alpha$ ($=0.057 \pm 0.002$ [GHz cm$^2$/W]) are fitting parameters.
The decay time constants of $\tau _{\textrm {x}}$ (=1.60 ns) and $\tau _{\textrm {xx}}$ (=0.72 ns) are evaluated by
the time-resolved PL measurements of $X$ and the corresponding biexciton state.
The above equations are derived from a three-level model,~\cite{Munsch09} which is constructed from
the simultaneous rate equations of the vacuum, exciton, and biexciton states. Details are shown in the Appendix.
In the range $P\ge P_{sat}$, $N(P)$ decreases indicating that the repumping process from the exciton to biexciton states is enhanced,
corresponding to the growth of additional peaks seen in Figs.~\ref{fig2}(b) and (c).
The side peaks of the $X$ line labeled as P1 and P2 are the biexciton and the excitonic complex~\cite{Finley01}
(EXC), respectively.
Approximately 50 meV higher energy emissions around 1 $\mathrm{\mu m}$ are attributed to EXCs related to the first excited state.
Cross-correlation measurements imply that all peaks appearing in Fig.~\ref{fig2}(b) originate from the same QD
showing dip and bunching behavior at zero time delay.

\begin{figure}[t]
\begin{center}
\includegraphics[width=20.5pc]{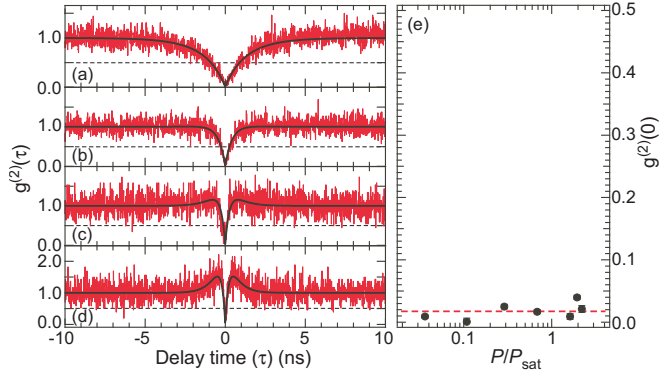}\hspace{2pc}
\caption{\label{fig4} Second-order photon correlation is recorded using a single photon counting module with two SNSPDs.
Normalized histograms of the auto-correlation measurement of the $X$ line below and above $P_{sat}$ are shown.
The solid red lines (a), (b), (c) and (d) are the raw data at $P/P_{sat}$ values of 0.036, 0.686, 1.625
and 2.202, respectively.
The solid black lines are fits taking the repumping process based on the three level system into account.
The excitation wavelength is 785 nm and the time bins are 12.2 ps. 
(e) $P/P_{sat}$ dependence of $g^{(2)}(0)$ with an error deduced by the standard deviation.
The dashed red line is the average $g^{(2)}(0)$ of 0.0174. }
\end{center}
\end{figure}

To develop this device into an SPS suitable for a fiber-based quantum-information network,
the high purity as a single photon is essential.
Therefore, we perform second-order photon correlation measurements to understand the single-photon nature of the $X$ line.
Measurements are performed  under non-resonant cw excitation conditions using a time-correlated single-photon-counting module
(Becker \& Hickl: SPC-130EM) and a pair of SNSPDs. The output of the device is filtered by a 0.5-nm-wide bandpass filter
to select the $X$ line and by two 50-nm-wide bandpass filters (Edmund Opt.: \#85-893) to suppress background photons, i.e.,
reflection of the excitation laser at the interface of the SMF patch cables and unwanted emissions originating
from the GaAs substrate.
Auto-correlation was measured by using a fiber-based HBT setup with a 50/50 fiber splitter and two SNSPDs (Fig.~\ref{fig1}(c)). 
Figure~\ref{fig4}(a) shows a histogram of the normalized coincidence with time bins of 12.2 ps at low excitation power
of $P/P_{sat}=0.036$. The data exhibits the well-known antibunching dip at zero time delay.
Below the saturation condition, this antibunching dip at zero time delay is clearly observed
even though the excitation power is increased (Fig.~\ref{fig4}(b)).
By increasing excitation power beyond the saturation condition,
a narrow antibunching structure remains, and a bunching extending over $\pm$2 ns near the narrow antibunching dip
appears as shown in Fig.~\ref{fig4}(c) and (d).
This bunching behavior originating from repumping to the biexciton state \cite{PRL87_257401} prevents us
from deducing $g^{(2)}(0)$ by using a simple two-level model.
To evaluate $g^{(2)}(0)$ of the $X$ line beyond $P_{sat}$, we use a correlation function based on the closed three-level model
(Appendix) shown below,
\begin {eqnarray}
g^{(2)}(\tau)&=&n_{\textrm{x}}(\tau)\frac {1-g^{(2)}(0)}{n_{\textrm{x}}(\infty)}+g^{(2)}(0), \label {eq7}\\
n_{\textrm{x}}(\infty)&=&\left ( 1+\frac {1}{\gamma \tau _{\textrm {x}}}+\gamma \tau _{\textrm{xx}} \right ) ^{-1}\label {eq8},
\end {eqnarray}
where $n_{\textrm {x}}(\tau)$ is the population of the exciton state.
The pumping rate $\gamma$ and $g^{(2)}(0)$ are fitting parameters. 
The solid curves in Figs.~\ref{fig4}(a)-(d)
are fitting results with $g^{(2)}(0)$ values of $0.0093 \pm 0.0027$, $0.0168 \pm 0.0032$, $0.0091 \pm 0.0045$ and $0.0212 \pm 0.0053$, and $\gamma$ values of
$0.034 \pm 0.001$ GHz, $0.639 \pm 0.026$ GHz, $1.514 \pm 0.060$ GHz and $2.052 \pm 0.082$ GHz
at $P/P_{sat}$ values of 0.036, 0.686, 1.625 and 2.202, respectively.
Figure~\ref{fig4}(e) shows the $P/P_{sat}$ dependence of $g^{(2)}(0)$ evaluated by Eqs.(\ref{eq7}) and (\ref{eq8}).
$g^{(2)}(0)$ does not change with increasing $P/P_{sat}$.
The average value of $g^{(2)}(0)$ is 0.0174 in the measured excitation power range,
signifying that background photon emission lowering the purity of single-photon nature is strongly suppressed in this device.
This means that the $X$ line persists in the pure single-photon state, even though the $X$ line saturates
and higher state emissions become prominent. 

\begin{figure}[t]
\begin{center}
\includegraphics[width=20.5pc]{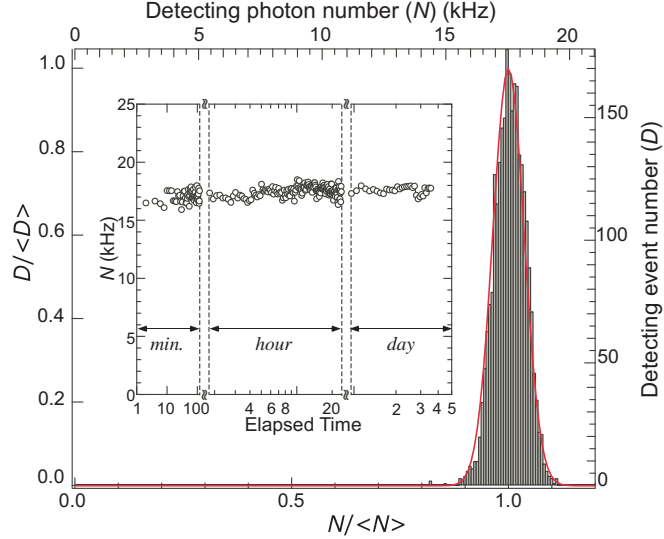}\hspace{2pc}%
\caption{\label{fig3} Long-term stability of $X$ line. Inset: Stability check for detecting photon count rate $N$ of the $X$ line
over four days. The red line is fitting curve with normal distribution function.}
\end {center}
\end{figure}

When discussing SPSs in a practical system, the stability of the emission wavelength and intensity also becomes a major issue.
Our SPS is directly coupled to the fiber end surface. The present device is mechanically solid, contrary to
the free-space optical setup which requires delicate treatment during the operation to sustain
the experimental condition against external disturbances.
To confirm the long-term stability of our SPS, we record the photon count rate $N$ of the $X$ line continuously
over four days at 2 minute intervals under the fixed excitation condition of $P_{sat}$.
The emission of the $X$ line passes through a 0.5-nm window and is sent to the SNSPD. 
The inset of Fig.~\ref{fig3} shows the variation of $N$ as a function of elapsed time.
The displayed data in the \textit{hour} and \textit{day} ranges are averaged over 8 minutes and 2 hours, respectively.
Figure~\ref{fig3} shows the frequency histogram of detected event number $D$ with each bin representing a 0.1-kHz period.
The bottom and left-hand axises are normalized by averaged values of $N$ and $D$, respectively.
The detected photon number is almost unchanged over four days, implying that the energy instability of the $X$ line is
lower than 0.6 meV, which corresponds to the 0.5-nm window of the bandpass filter.
The fluctuation $(1\sigma )$ is $3.61 \pm 0.03 \%$ which is deduced by fitting with a normal distribution function
(red curve in Fig.~\ref{fig3}), suggesting that the present device has large potential as a photon source
with the stability of photon number. It is worth noting that $N$ remains almost unchanged after repeating the measurement
over several months of heat cycles forced by changes of the liquid-He vessel.
This means that this sample is quite rigid against external disturbances such as cooling-heating cycles and vibration. 

In addition to {\it purity} and {\it stability}, {\it efficiency} is also a key factor for practical applications.
Here we estimate the coupling efficiency from the SPS pillar array to the SMF. In the case where the $X$ line is excited
by the saturation power $P_{sat}$, photons are filled up with a time interval of $\tau_{X}$.
Therefore, the generated photon number at $P_{sat}$ is $\sim 1/\tau_{X} =0.625$ GHz.
On the other hand, the observed photon number by two SNSPDs at $P_{sat}$ is $2 \times N(P_{sat}) = 25.0$ kHz.
Considering that the quantum efficiency of SNSPD is $\sim 16 \%$ and the total throughput of our optical setup is $\sim 9.8 \%$,
we can conclude that the coupling efficiency from the SPS pillar array to the SMF is $\sim 0.26 \%$.
The present sample works as an SPS with high purity and stability, but further improvement is necessary for efficiency.
One way of improving the efficiency is positional matching between the pillar and the fiber core center.
The use of a fiber-coupled SPS module that can optimize the positional relationship between the fiber and the SPS
by using a piezo positioner \cite{APL99_121101, SR5_14309, APEX9_032801} is going in this direction.
Another way is to improve the photon extraction efficiency from pillars. The metal-embedded SPS
\cite{APEX6_062801, APL102_131114, JAP116_043103} succeeded in improving the extraction efficiency by up to
18\% , and by up to 24.6\% with a cone structure.
It is worth noting that the width of the X line, 0.6 meV is comparable to the window of the bandpass filter.
Therefore the side part of the X line is possibly cut off, then the photon count rate is underestimated.
The indistinguishability of emitted photons from single QD is an another essential aspect for the future applications.
QDs in nanostructures are susceptible to proximity effects inducing linewidth broadening, \cite{APL85_3423}
as well as short timescale ($< 100$ ms) photon emission intermittency. \cite{PRB89_161303(R)}
In order to suppress these phenomena, the QD should be located sufficiently far from any surfaces.
In our case, the distance from the sidewall of a pillar to the QD is crucial.
Uniformity of the emission energy of the QD ensemble is also important ingredient to endorse the practical yield
of indistinguishability.
In-flush technique \cite{JCG201-202_1131, JAP102_013515} or
thermal annealing \cite{PRB69_161301R, APL89_263109, APL90_011907} will be effective to this direction,
and the external electrical \cite{APL86_041907, APL90_041101}, magnetic \cite{PRB73_033306},
mechanical \cite{APL88_203113} field and their combination will be useful to tune the emission energy and broadening.
Optimizing the pillar array SPS with these foresights, we are planning to introduce fiber bundle \cite{APL93_021124}
to realize indistinguishability by selecting highly identical emission lines.

\section{Conclusion}

An InAs QD pillar array SPS was fabricated. Pillars, one-at-a-time, were directly coupled to the fiber core of an SMF.
The pillar diameter and distance between pillars were well controlled as only a few QDs could couple
to the fiber. We performed an auto-correlation measurement on a well-defined emission peak using the HBT setup.
Our results showed an average $g^{(2)}(0)$ value of 0.0174 over the measured excitation power range.
This pillar array had a mechanically solid structure.
We performed photon counting experiments at fixed wavelength with a 0.5-nm window
continuously over several days. Our results showed long-term stability against external disturbances such as cooling-heating
cycles and vibration. We believe that our sample provides high quality photons that enables the quantum information
technology in practical optical-fiber networks.



%
%

%

\begin{acknowledgments}
This work was partly supported by Strategic Information and Communications R\&D Promotion Programme (SCOPE),
JSPS KAKENHI Grant Number 16H03816, 16H03817, 17K06396,
and Cooperative Research Program of ``Network Joint Research Center for Materials and Devices''.
\end{acknowledgments}

\section* {Appendix: Three-level rate equation}

\setcounter{equation}{0}
\def\theequation{A\arabic{equation}}

In the weak excitation case, we usually use the simple two-level model. However, as the excitation power increases,
the exciton state is filled and the biexciton correction becomes necessary.
Here we introduce the simultaneous rate equations of the vacuum, exciton, and biexciton states,
\begin {eqnarray}
\frac{dn_{0}}{dt}&=&-\gamma n_{0}+\frac{n_{\textrm{x}}}{\tau_{\textrm{x}}},\label{appendix-eq01} \\
\frac {dn_{\textrm {x}}}{dt}&=&-\frac{n_{\textrm{x}}}{\tau_{\textrm{x}}}+\frac {n_{\textrm{xx}}}{\tau_{\textrm{xx}}}
  +\gamma n_{0} -\gamma^{\prime} n_{\textrm{x}},\label{appendix-eq02} \\
\frac {dn_{\textrm {xx}}}{dt}&=&-\frac {n_{\textrm {xx}}}{\tau_{\textrm{xx}}}+\gamma^{\prime} n_{\textrm{x}}\label{appendix-eq03},
\end {eqnarray}
where $n_{0}$, $n_{\textrm{x}}$ and $n_{\textrm{xx}}$ are the occupation numbers of each state,
$\gamma$ and $\gamma^{\prime}$ are the pumping rates from the vacuum to exciton and from the exciton to biexciton states,
respectively.
$n_{\textrm{x}}$ is the sum of two exciton states by spin degrees of freedom.
With the constraint
\begin{equation}
n_0 +n_{\textrm {x}}+n_{\textrm {xx}}=1,\label{appendix-eq04}
\end{equation}
we can rewrite Eq. (\ref{appendix-eq02}) as
\begin{eqnarray}
\frac{dn_{\textrm {x}}}{dt}&=&\frac{1}{\tau_{\textrm{xx}}}+\left( \gamma-\frac{1}{\tau_{\textrm{xx}}} \right) n_{0} \nonumber \\
&-&\left( \frac{1}{\tau_{\textrm{x}}}+\frac{1}{\tau_{\textrm{xx}}}+\gamma^{\prime} \right) n_{\textrm{x}}.\label{appendix-eq05}
\end{eqnarray}
Deriving second-order differential equations from Eqs. (\ref{appendix-eq01}) and (\ref{appendix-eq05}), we can separate
$n_{0}$ and $n_{\textrm{x}}$ as below,
\begin{eqnarray}
\frac{d^{2}n_{0}}{dt^{2}}
&+&\left( \frac{1}{\tau_{\textrm{x}}}+\frac{1}{\tau_{\textrm{xx}}}+\gamma +\gamma^{\prime }\right)
\frac{dn_{0}}{dt} \nonumber \\
&+&\left( \frac{1}{\tau_{\textrm{x}}\tau_{\textrm{xx}}}+\frac{\gamma}{\tau_{\textrm{xx}}}+\gamma \gamma^{\prime} \right)
n_{0} =\frac{1}{\tau_{\textrm{x}}\tau_{\textrm{xx}}}, \label{appendix-eq06} \\
\frac{d^{2}n_{\textrm{x}}}{dt^{2}}
&+&\left( \frac{1}{\tau_{\textrm{x}}}+\frac{1}{\tau_{\textrm{xx}}}+\gamma +\gamma^{\prime}\right)
\frac{dn_{\textrm{x}}}{dt} \nonumber \\
&+&\left( \frac{1}{\tau_{\textrm{x}}\tau_{\textrm{xx}}}+\frac{\gamma}{\tau_{\textrm{xx}}}+\gamma \gamma^{\prime} \right)
n_{\textrm{x}} =\frac{\gamma}{\tau_{\textrm{xx}}}. \label{appendix-eq07}
\end{eqnarray}
We obtain
\begin{eqnarray}
n_{0}(t)&=&N^{(+)}_{0}\exp (-\alpha_{+}t) +N^{(-)}_{0}\exp (-\alpha_{-}t) \nonumber \\
&+&\left( 1+\gamma \tau_{\textrm{x}}+\gamma \gamma^{\prime} \tau_{\textrm{x}}\tau_{\textrm{xx}}
\right)^{-1}, \label{appendix-eq08} \\
n_{\textrm{x}}(t)&=&N^{(+)}_{\textrm{x}}\exp(-\alpha_{+}t) +N^{(-)}_{\textrm{x}}\exp (-\alpha_{-}t) \nonumber \\
&+&\left( 1+\frac{1}{\gamma \tau_{\textrm{x}}}+\gamma^{\prime} \tau_{\textrm{xx}}\right)^{-1}. \label{appendix-eq09}
\end{eqnarray}
$\alpha_{+}$ and $\alpha_{-}$ are derived from characteristic equations of (\ref{appendix-eq06})
and (\ref{appendix-eq07}),
\begin{eqnarray}
\alpha_{\pm} &=& \frac{ \displaystyle \frac{1}{\tau_{\textrm{x}}}+\frac{1}{\tau_{\textrm{xx}}}+\gamma +\gamma^{\prime}
\pm \sqrt{ \Delta^{2}+\frac{4\gamma^{\prime}}{\tau_{\textrm{x}}}}}{2}, \label{appendix-eq10} \\
\Delta &=& -\frac{1}{\tau_{\textrm{x}}}+\frac{1}{\tau_{\textrm{xx}}} -\gamma +\gamma^{\prime}. \label{appendix-eq11}
\end{eqnarray}
It is important that Eqs. (\ref{appendix-eq08}) and (\ref{appendix-eq09}) have two time constants, $\alpha_{+}$ and $\alpha_{-}$,
with exponential decay. Considering $\Delta^{2} \gg 4\gamma^{\prime}/\tau_{\textrm{x}}$, we have
$\alpha_{+} \simeq 1/\tau_{\textrm{xx}}+\gamma^{\prime}$ and $\alpha_{-} \simeq 1/\tau_{\textrm{x}}+\gamma$.
These two different time components provide the dip structure to $n_{\textrm{x}}(t)$.
This is the main difference from the result for the two-level model with only one component of exponential decay.

Now, we evaluate the auto-correlation of the excitons.
Suppose that we have two single-photon detectors $D_{1}$ and $D_{2}$. By using $D_1$ as the ``start'' of photon counting,
$n_{\textrm{x}}$ of $D_{2}$ at relative time $\tau$ gives the correlation between $D_1$ and $D_2$ at $\tau$.
Including the accidental coincidence $B$, $g^{(2)}(\tau) \propto B+An_{\textrm{x}}(\tau)$.
Therefore, normalizing it by the value at $\tau = \infty$, we have
\begin{eqnarray}
g^{(2)}(\tau) = \frac{B+An_{\textrm{x}}(\tau)}{B+An_{\textrm{x}}(\infty)}. \label{appendix-eq12}
\end{eqnarray}
With the initial condition $n_{\textrm{x}}(0) =0$, we have
\begin{eqnarray}
g^{(2)}(0)=\frac{B}{B+An_{\textrm{x}}(\infty)}. \label{appendix-eq13}
\end{eqnarray}
Equations (\ref{appendix-eq12}) and (\ref{appendix-eq13}) yield Eq. (\ref{eq7}) in the main text.
From the abovementioned maesurement procedure by $D_1$ and $D_2$,
we can determine that $g^{(2)}(\tau)$ is an even function.
Therefore, we redefine $n_{\textrm{x}}$ as
\begin{eqnarray}
n_{\textrm{x}}(\tau)&=&N^{(+)}_{\textrm{x}}\exp(-\alpha_{+}\left| \tau \right| )
 +N^{(-)}_{\textrm{x}}\exp (-\alpha_{-}\left| \tau \right| ) \nonumber \\
&+&\left( 1+\frac{1}{\gamma \tau_{\textrm{x}}}+\gamma^{\prime}\tau_{\textrm{xx}}\right)^{-1}. \label{appendix-eq14}
\end{eqnarray}
Equation (\ref{eq8}) is derived from this equation at $\tau = \infty$ with $\gamma^{\prime}=\gamma$.
The pumping rate is proportional to the excitation power $P$.
Therefore, setting $\gamma =\alpha P$ in Eq. (\ref{eq8}), we have Eq. (\ref{eq1}).
It is worth noting that we can obtain Eq. (\ref{eq8}) simply by forcing $d^2n_{\textrm{x}}/dt^2=dn_{\textrm{x}}/dt =0$
in Eq. (\ref{appendix-eq07}).

\begin{figure}[t]
\begin{center}
\includegraphics[width=20.5pc]{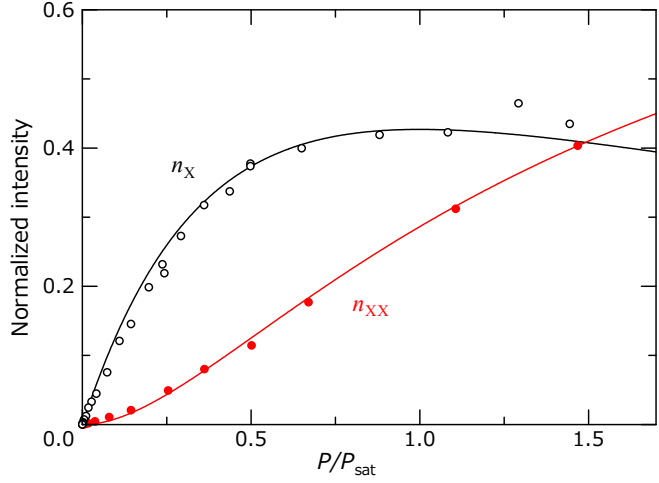}\hspace{2pc}%
\caption{\label{fig-appendix} $P$ dependence of $n_{\textrm{x}}(\infty)$ (black line)
 and $n_{\textrm{xx}}(\infty)$ (red line).
Corresponding exciton $X$ (black open circle) and biexciton P1 (red filled circle) normalized intensity in Fig.\ref{fig2}(b)
are plotted for comparison. Parameters $\tau_{\textrm{x}}$, $\tau_{\textrm{xx}}$ and $\alpha$ are shown in the main text.}
\end {center}
\end{figure}

In the last, we give an overview of the correspondence between results by this three-level model and experimental ones.
$n_{0}(\infty)$ and $n_{\textrm{x}}(\infty)$ are given by
\begin{eqnarray}
n_{0}(\infty)&=&
 \frac{1}{1+\alpha P \tau_{\textrm{x}}+\alpha^2 P^2\tau_{\textrm{x}}\tau_{\textrm{xx}}}, \\
n_{\textrm{x}}(\infty)&=&\
\frac{\alpha P\tau_{\textrm{x}}}
 {1+\alpha P \tau_{\textrm{x}}+\alpha^2 P^2\tau_{\textrm{x}}\tau_{\textrm{xx}}}.
\end{eqnarray}
With the constraint of Eq. (\ref{appendix-eq04}), we have
\begin{eqnarray}
n_{\textrm{xx}}(\infty)&=&
\frac{\alpha^2 P^2\tau_{\textrm{x}}\tau_{\textrm{xx}}}
{1+\alpha P\tau_{\textrm{x}}+\alpha^2 P^2\tau_{\textrm{x}}\tau_{\textrm{xx}}}.
\end{eqnarray}
As shown in Fig. \ref{fig-appendix}, we have reasonable correspondence between model calculation and experiments
in the wide excitation power $P$ range even if $P$ is beyond the saturation $P_{sat}$.


\bibliographystyle{aipnum4-1}

\bibliography{reference_odashima}

\end{document}